# Fast long-wavelength exchange spin waves in partially-compensated Ga:YIG

T. Böttcher,[1, 2] M. Ruhwedel,[1] K. O. Levchenko,[3] Q. Wang,[3] H. L. Chumak,[4] M. A. Popov,[4] I. V. Zavislyak,[4] C. Dubs,[5] O. Surzhenko,[5] B. Hillebrands,[1] A. V. Chumak,[3] and P. Pirro[1]

[1)]*Fachbereich Physik and Landesforschungszentrum OPTIMAS, Technische Universität Kaiserslautern, Gottlieb-Daimler-Straße 46, 67663 Kaiserslautern, Germany*
[2)]*MAINZ Graduate School of Excellence, Staudingerweg 9, 55128 Mainz, Germany*
[3)]*Faculty of Physics, University of Vienna, A-1090 Wien, Austria*
[4)]*Faculty of Radiophysics, Electronics and Computer Systems, Taras Shevchenko National University of Kyiv, Kyiv, 01601, Ukraine*
[5)]*INNOVENT e.V. Technologieentwicklung, Prüssingstrasse 27B, 07745 Jena, Germany*

(Dated: 20 December 2021)

Spin waves in yttrium iron garnet (YIG) nano-structures attract increasing attention from the perspective of novel magnon-based data processing applications. For short wavelengths needed in small-scale devices, the group velocity is directly proportional to the spin-wave exchange stiffness constant $\lambda_{\text{ex}}$. Using wave vector resolved Brillouin Light Scattering (BLS) spectroscopy, we directly measure $\lambda_{\text{ex}}$ in Ga-substituted YIG thin films and show that it is about three times larger than for pure YIG. Consequently, the spin-wave group velocity overcomes the one in pure YIG for wavenumbers $k > 4$ rad/$\mu$m, and the ratio between the velocities reaches a constant value of around 3.4 for all $k > 20$ rad/$\mu$m. As revealed by vibrating-sample magnetometry (VSM) and ferromagnetic resonance (FMR) spectroscopy, Ga:YIG films with thicknesses down to 59 nm have a low Gilbert damping ($\alpha < 10^{-3}$), a decreased saturation magnetization $\mu_0 M_S \approx 20$ mT and a pronounced out-of-plane uniaxial anisotropy of about $\mu_0 H_{\text{u1}} \approx 95$ mT which leads to an out-of-plane easy axis. Thus, Ga:YIG opens access to fast and isotropic spin-wave transport for all wavelengths in nano-scale systems independently of dipolar effects.

Wave-based logic concepts[1-3] are expected to come along with major advantages over information processing based on current CMOS-based information technology[4,5]. In particular, coherent spin waves[6] are envisaged to allow for the realization of efficient wave-based logic devices[3, 7-12]. However, progress in this field places high demands on the materials used. Specifically, spin-wave elements operating at large clock frequencies demand materials which exhibit a small Gilbert damping constant[13], large spin-wave velocities and good processing properties. Yttrium Iron Garnet (YIG) has a very small damping constant[14] and nano-structures of 50 nm lateral sizes have been demonstrated recently[15-17]. However, dipolar spin waves in YIG waveguides feature velocities that are significantly reduced compared to plane YIG films[16] which is caused by the flattening of the dispersion curve. The fastest dipolar waves in these structures are magnetostatic surface waves which offer a group velocity of around 0.2 $\mu$m/ns[17]. Exchange spin waves with wavelengths in the range of 100 nm or shorter are faster[18] but the excitation of such short wavelengths is a separate challenge[16]. It can be addressed by the utilization of: nanoscopic antennas with increased Ohmic loss; strongly non-uniform magnetic patterns[19]; hybrid nanostructures utilizing different magnetic materials[20-22]; or more complex physical phenomena like magnon Cherenkov radiation[23].

To operate with waves of maximized speed in nano-structures, materials with large spin-wave exchange stiffness $\lambda_{\text{ex}} = 2A/(\mu_0 M_S)$ are mandatory since the exchange contribution to the group velocity is directly proportional to $\lambda_{\text{ex}}$. Here, $A$ is the Heisenberg exchange constant, $M_S$ the saturation magnetization and $\mu_0$ is the permeability of the vacuum. In addition, the possibility to operate with fast exchange-dominated spin waves of larger wavelengths would not only allow for the operations with "standard" micro-scaled antennas[11,16], but also would give the freedom required for the engineering of data-processing units[1,24] since the exchange-dominated dispersion relation is highly isotropic.

In this context, it is promising to study ferrimagnetic insulators which are close to the magnetic compensation since low $M_S$ tends to increase $\lambda_{\text{ex}}$. For this, Liquid Phase Epitaxy (LPE) based[14] growth of Ga-substituted YIG[25-28], which has been adopted for the deposition of $Y_3Fe_{5-x}Ga_xO_{12}$ single crystalline films of sub-100 nm thicknesses, is very interesting. In these films, non-magnetic $Ga^{3+}$ ions preferentially substitute the magnetic $Fe^{3+}$ ions in the tetrahedral coordinated magnetic sub-lattice ($0 < x < 1.5$), decreasing the $M_S$ of this ferrimagnetic material down to the fully compensated antiferromagnetic state (for the single crystals grown from the high-temperature solutions at contents $x$ of about 1.27 formula units at room temperature $T = 295K$)[25,26]. Besides, the Ga substitution induces a strong out-of-plane uniaxial anisotropy enabling the easy axis of the thin film perpendicular to the film surface.

Here, we report on the investigation of the spin waves in Ga:YIG films of sub-100 nm thickness. Ga substitution $x \approx 1$ was chosen so that the magnetization of the sample is decreased to one tenth of the pure YIG value. This was done in order to change the easy magnetization orientation from the in-plane to out-of-plane, simultaneously avoiding an excessive increase in the magnetic damping. First, a thorough characterization is performed using ferromagnetic resonance spectroscopy (FMR) in combination with vibrating-sample magnetometry (VSM) to obtain the saturation magnetization, the anisotropy constants and the damping parameters. Afterward, wave vector resolved[29] Brillouin light scattering (BLS) spectroscopy is used to measure the dispersion relation of thermal spin waves $\omega(k)$ and the spin-wave exchange stiffness $\lambda_{ex}$ directly. The resulting group velocity $d\omega/dk$ is compared to the velocities of spin waves in pure YIG films. Our presented results are complementary and in good agreement with recent results from an indirect measurement of the dispersion relation at low wave vectors using electrical spectroscopy in Ga:YIG reported by Carmiggelt et al.[30]

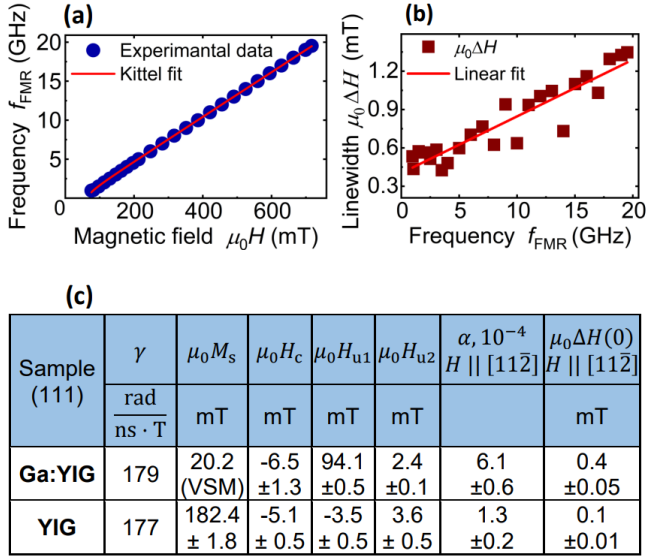

FIG. 1. (a) Ferromagnetic resonance frequency $f_{FMR}$ as a function of the in-plane magnetic field $H \parallel [11\bar{2}]$. Experimental data (blue dots) is fitted by a modified Kittel equation (solid red line). (b) Recalculated full width at half maximum (FWHM) field linewidth $\mu_0 \Delta H$ as a function of $f_{FMR}$. Measured data (dark red squares) is fitted by a linear regression (solid red line). A full evaluation (see Supplementary Materials) provides the values shown in the Table (c). (c) Parameters obtained from FMR-VNA data of 59 nm Ga:YIG and 97 nm YIG films grown on GGG(111).

In the following, we present exemplary Broadband Ferromagnetic Resonance – Vector Network Analyser (FMR - VNA) spectroscopy data for a LPE-grown 59 nm thick film of Ga:YIG/GGG(111). Data for the other films thicknesses of 105 nm-thick Ga:YIG/GGG(111), 95 nm-thick Ga:YIG/GGG(001), and a reference film of 97 nm-thick YIG/GGG(111) are provided in the Supplementary Materials. The films grown on GGG(111) were cut into the square specimens with the edges oriented along $[11\bar{2}]$ and $[1\bar{1}0]$ crystallographic directions, and samples on GGG (001) – along [100] and [010] (see Figs. S1 and S5 in the Supplementary Materials). For magnetic characterization, FMR-VNA was performed in the frequency range up to 20 GHz and at the rf power of 0 dBm. To define the crystallographic parameters of the samples, theoretical model of Bobkov and Zavislyak was used[31]. The model differentiates three main anisotropy fields – a cubic field $H_c$, a uniaxial anisotropy field of the first order $H_{u1}$, and a uniaxial anisotropy field of the second order $H_{u2}$. The cubic anisotropy originates from the magnetization along the preferred crystallographic directions in the garnet lattice[14,31], while the uniaxial anisotropy consists of cubic, growth-induced and strain-induced contributions resulting in an effective uniaxial anisotropy[14,32,33]. In addition to the direction of a magnetic field $H$, the FMR frequency also depends on the crystallographic orientation of the GGG substrate[31]. To define all the anisotropy fields experimentally, a magnetic field was applied in-plane (IP) along the two orthogonal crystallographic axes $[11\bar{2}]$ and $[1\bar{1}0]$ (corresponding to the sample's edges), and out-of-plane (OOP) along the [111] direction normal to the film plane. The saturation magnetization $M_S$ was measured by VSM. Detailed descriptions of the employed theoretical model, the measurement procedure, and the FMR analyses for the film under the investigation are given in the Supplementary Materials.

The dependence of the FMR frequency on the in-plane magnetic field is shown in Fig. 1(a). The Gilbert damping parameter $\alpha$ and the inhomogeneous linewidth broadening $\Delta H(0)$ were found according to the standard approach described in Ref.[34] (see Fig. 1(b)), taking into consideration that $H_{u1} > M_S$. All obtained parameters are summarized in Fig. 1(c) along with the results of a reference 97 nm-thick YIG/GGG(111) film. A comparison of these values demonstrates that the Ga doping led to a ∼ 9 times reduction in the saturation magnetization $M_S$, a ∼ 27 times increase in the uniaxial anisotropy $H_{u1}$, and a ∼ 4.7 times increase in the Gilbert damping $\alpha$. A strong increase of the uniaxial anisotropy indicates the out-of-plane easy axis of Ga:YIG thin film. While an increase in $\alpha$ was observed, it is still significantly lower compared to metallic compounds[2]. The obtained values are in good agreement with those reported in Ref[30].

After the characterization via FMR, the dispersion relation of thermally excited, magnetostatic surface spin waves propagating perpendicularly to the applied field is probed by Brillouin light scattering spectroscopy (BLS)[29] to obtain the exchange stiffness $\lambda_{ex}$. An external field of $\mu_0 H = 300$ mT is applied in the film plane along the sample edge which ensures an in-plane magnetization. For the spectral analysis of the scattered light, a 6-pass tandem Fabry-Pérot interferometer is used[35]. In all measurements presented here, a laser with a wavelength of $\lambda_{Laser} =$

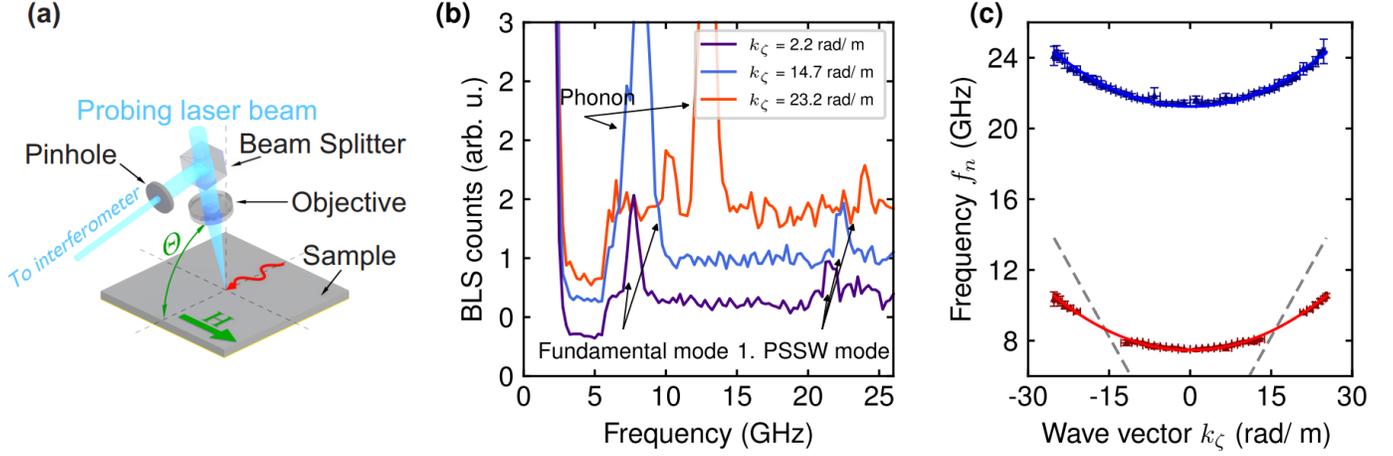

FIG. 2. (a) Schematic BLS setup. (b) Three exemplary BLS spectra obtained for the different in-plane wave vectors from a 59 nm thick Ga:YIG film at an applied field of 300 mT. Two spin-wave modes as well as a phonon mode can be observed. (c) Dispersion relations for the spin-wave modes extracted from all the measured BLS spectra. The solid lines are fits according to the model given in Eqn.1. The dashed lines are linear fits of the phonon mode.

491 nm is used. The in-plane component $k_\zeta$ of the wave vector of the probed spin wave is varied by changing the angle of light incidence $\Theta$, $k_\zeta = 4\pi \sin(\Theta)/\lambda_{\text{Laser}}$ — see Fig. 2(a). The results for the 59 nm thick Ga:YIG film are presented in Fig. 2. Fig. 2(b) shows three exemplary BLS spectra (anti-stokes part). Data for the three different in-plane wave vectors are presented. Besides the quite strong phonon signal, one can distinguish the fundamental spin-wave mode and the first perpendicular standing spin-wave mode (PSSW). For wave vectors between 12 rad µm$^{-1}$ and 20 rad µm$^{-1}$ no fundamental mode could be observed because of the strong signal of the phonon mode that crosses the fundamental spin-wave mode in this area (see, e.g., the spectrum for $k_\zeta = 14.7$ rad µm$^{-1}$).

The corresponding analytical description of the spin-wave dispersion relation for the case of a ferromagnetic film in (111) orientation having uniaxial and cubic anisotropy with unpinned surface spins has been obtained by Kalinikos et al.[36]. The wave vector quantization along the film normal, which results in the appearance of the PSSWs[37], is described by the index $n$ such that the dispersion relation is given by:

$$f_n(\mathbf{k}) = \frac{\gamma\mu_0}{2\pi}\sqrt{\left(H + \lambda_{\text{ex}}k_n^2 + M_S - M_S P_{nn}(k_\zeta t) - H_c - H_{u1}\right)\left(H + \lambda_{\text{ex}}k_n^2 + M_S P_{nn}(k_\zeta t)\sin^2\phi\right) - 2H_c^2\cos^2 3\phi_M} \quad (1)$$

where $k_n = \sqrt{k_\zeta^2 + \kappa_n^2}$ is the total spin wave vector consisting of the in-plane spin wave vector $k_\zeta$ and the out-of-plane spin wave vector $\kappa_n$ with $\kappa_n = \frac{n\pi}{t}$, $n = 0, 1, 2, ...$ Here, $t$ is the thickness of the film, $\phi$ is the angle between the static magnetization and the in-plane wave vector $k_\zeta$, $\phi_M$ is the angle between the static magnetization and the $[1\bar{1}0]$ axis, $H$ is the applied magnetic field and $\gamma$ is the gyromagnetic ratio which we take from the FMR measurements. The matrix element $P_{nn}$ is a function of $k_\zeta t$ ($0 \leq P_{nn} < 1$ if $0 \leq k_\zeta t < \infty$). In the long wavelength limit ($k_\zeta t \ll 1$) and for unpinned surface spins the following approximations have been obtained: $P_{00} = \frac{k_\zeta t}{2}$ for $n = 0$ and $\left(\frac{k_\zeta t}{n\pi}\right)^2$ for $n \neq 0$[38].

The dispersion relations of the respective modes extracted from the BLS spectra are shown in Fig. 2(c) together with fit curves according to Eqn. 1. FMR measurements show that $H_c$ is about one order of magnitude smaller than $H_{u1}$ (compare Fig. 1(c)). Consequently, the last term in Eqn.1 that is quadratic in $H_c$ can be safely neglected. For the fits we have fixed the saturation magnetization to $\mu_0 M_S = 20.2$ mT as obtained from VSM, and the gyromagnetic ratio to $\gamma = 179$ rad T$^{-1}$ns$^{-1}$ as obtained from the FMR measurements. The extracted values from the simultaneous fits of the fundamental mode and the first PSSW mode are: exchange stiffness $\lambda_{\text{ex}} = (13.54 \pm 0.07) \times 10^{-11}$ Tm$^2$, exchange constant $A = (1.37 \pm 0.01)$ pJm$^{-1}$, respectively. The sum of the anisotropy fields is $\mu_0(H_u + H_c) = (91.3 \pm 0.4)$ mT, in very good agreement with the values obtained from FMR (compare to Fig. 1(c)).

The exchange stiffness in the film under investigation in this work is about three times as large as the one for pure

YIG[39]. This results in a much higher group velocity than in pure YIG as can be seen in Fig. 3. There the fitted dispersion relation of the fundamental mode for the investigated Ga:YIG film and the corresponding dispersion relation for a pure YIG film of the same thickness of 59 nm and at the same applied field of 300 mT is shown. Here the standard parameters of YIG[14,39] have been used: $\gamma = 177$ rad $T^{-1}ns^{-1}$ (from FMR), $\lambda_{ex} = 4.03 \cdot 10^{-11}$ Tm², $\mu_0 M_S = 177.2$ mT (all anisotropy contributions are neglected). The corresponding group velocities calculated by $v_{gr} = 2\pi \partial f_n(\mathbf{k})/\partial k_n$ are plotted in the lower part of Fig. 3(b). The exchange dominated region is characterized by a linear dependence of the group velocity on the wave vector. Thus, spin waves in the Ga:YIG can be considered as exchange dominated down to very low wave vectors as it is directly visible from Fig. 3(b). For wave vectors above $k > 4$ rad/µm, spin-waves in Ga:YIG are faster compared to pure YIG.

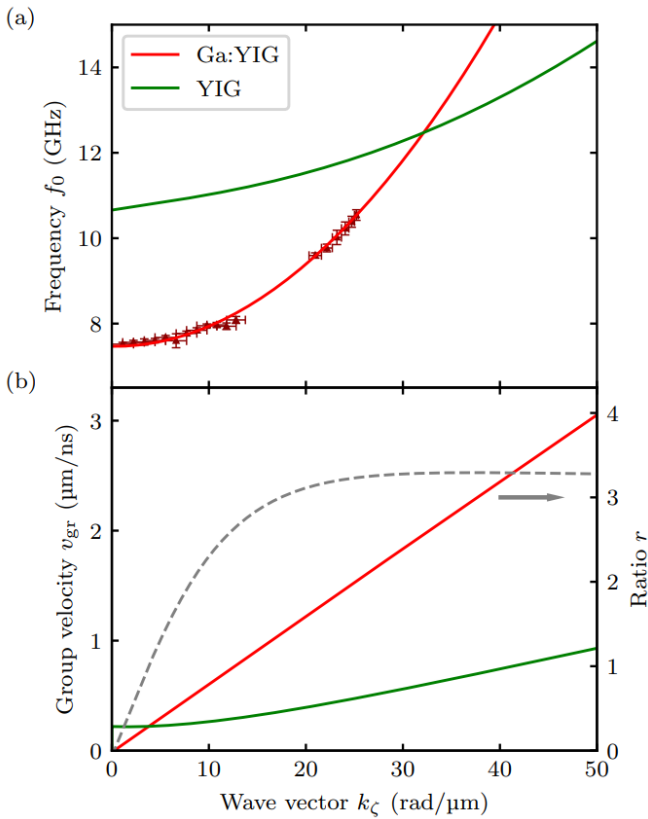

FIG. 3. (a) Dispersion fitted to the measured data from the investigated 59 nm thick Ga:YIG film at an applied field of 300 mT according to Eqn.1 (red) and a theoretical dispersion calculated according to Eqn.1 for a pure YIG film (green) of the same thickness at the same applied field using standard YIG parameters (see text)[14,39]. (b) Group velocity calculated from the dispersion relation in (a) for Ga:YIG (red) and pure YIG (green). The ratio $r$ of the group velocities for Ga:YIG and pure YIG is plotted by a grey dashed line.

For wave vectors $k > 30$ rad/µm, both dispersion relations are dominated and the ratio $r$ of the group velocities is converging to the ratio of the exchange stiffness constants $r \approx \lambda_{ex}(\text{Ga:YIG})/\lambda_{ex}(\text{YIG}) \approx 3.4$.

In conclusion, we have investigated spin-wave properties in Ga-substituted YIG with significantly decreased saturation magnetization $\mu_0 M_S \approx 20.2$ mT and increased exchange stiffness $\lambda_{ex} = (13.54 \pm 0.07) \cdot 10^{-11}$ Tm². The saturation magnetization $M_S$ was measured using VSM, the three anisotropy constants $H_c$, $H_{u1}$, $H_{u2}$ and the gyromagnetic ratio $\gamma$ were determined using FMR, and the exchange stiffness $\lambda_{ex}$ was determined from BLS measurements of the dispersion relation of the fundamental and the first PSSW mode. We find that even spin waves of relatively small wave vector $k \approx 4$ rad/µm exhibit an exchange nature, and their velocities are higher than in pure YIG, reaching a ratio of approximately 3.4 as defined by the ratio of the individual exchange stiffness constants. As a further consequence, waves in Ga:YIG have a significantly more isotropic dispersion relation than waves of the same wavelength in YIG. Thus, for magnonic waveguides structured from Ga:YIG, only a weak dependence of important parameters, such as the wave velocity and the wave phase accumulation, on the structure sizes and on the magnetization orientation can be expected.

The small saturation magnetization and the uniaxial anisotropy lead to an out-of-plane easy axis which facilitates also the use of the entirely isotropic Forward Volume waves. Since the relative drop of the exchange constant $A$ with Ga substitution $x$ is weaker than the drop of the saturation magnetization $M_S$, one can expect that a further reduction of $M_S$ by an increased Ga substitution will lead to even faster and more isotropic spin waves. Eventually, a fully compensated Ga:YIG film might serve as a model system for antiferromagnetic magnonics. Thus, Ga:YIG opens access to the operation with fast and isotropic exchange spin waves of variable wavelengths in future magnonics networks.

**ACKNOWLEDGMENTS**

This research has been funded by the Deutsche Forschungsgemeinschaft (DFG, German Research Foundation) - 271741898, by the DFG Collaborative Research Center SFB/TRR 173-268565370 (Projects B01 and B11), by the Austrian Science Fund (FWF) through the project I 4696-N, and by the European Research Council project ERC Starting Grant 678309 MagnonCircuits. The authors thank Volodymyr Golub (Institute of Magnetism, National Academy of Sciences of Ukraine) for support and valuable discussions, as well as M. Lindner and T. Reimann (INNOVENT e.V.) for the production of the YIG reference sample and R. Meyer for the technical assistance.

## DATA AVAILABILITY

The data that support the findings of this study are available from the corresponding author upon reasonable request.

## REFERENCES

[1] A. Mahmoud, F. Ciubotaru, F. Vanderveken, A. V. Chumak, S. Hamdioui, C. Adelmann, and S. Cotofana, J. Appl. Phys. **128**, 161101 (2020).

[2] A. Barman, G. Gubbiotti, S. Ladak, A. O. Adeyeye, M. Krawczyk, J. Gräfe, C. Adelmann, S. Cotofana, A. Naeemi, V. I. Vasyuchka, B. Hillebrands, et al., J. Phys.: Condens. Matter **33** (2021).

[3] A. V. Chumak, P. Kabos, M. Wu, C. Abert, C. Adelmann, A. Adeyeye, J. Åkerman, F. G. Aliev, A. Anane, A. Awad, C. H. Back, et al., (2021), arXiv:2111.00365.

[4] M. M. Waldrop, Nature **530**, 144 (2016).

[5] B. Dieny, I. L. Prejbeanu, K. Garello, P. Gambardella, P. Freitas, R. Lehndorff, W. Raberg, U. Ebels, S. O. Demokritov, J. Akerman, et al., Nat. Electron. **3**, 446 (2020).

[6] P. Pirro, V. Vasyuchka, A. A. Serga, and B. Hillebrands, Nat. Rev. Mater. **6**, 1114 (2021).

[7] A. Khitun, M. Bao, and K. L. Wang, J. Phys. D: Appl. Phys. **43**, 264005 (2010).

[8] A. V. Chumak, A. A. Serga, and B. Hillebrands, Nat. Commun. **5**, 4700 (2014).

[9] T. Fischer, M. Kewenig, D. A. Bozhko, A. A. Serga, I. I. Syvorotka, F. Ciubotaru, C. Adelmann, B. Hillebrands, and A. V. Chumak, Appl. Phys. Lett. **110**, 152401 (2017).

[10] G. Talmelli, T. Devolder, N. Träger, J. Förster, S. Wintz, M. Weigand, H. Stoll, M. Heyns, G. Schütz, I. P. Radu, et al., Sci. Adv. **6**, eabb4042 (2020).

[11] Q. Wang, M. Kewenig, M. Schneider, R. Verba, F. Kohl, B. Heinz, M. Geilen, M. Mohseni, B. Lägel, F. Ciubotaru, et al., Nat. Electron. **3**, 765 (2020).

[12] A. N. Mahmoud, F. Vanderveken, C. Adelmann, F. Ciubotaru, S. Hamdioui, and S. Cotofana, IEEE Trans. Magn. **57**, 1 (2021).

[13] T. L. Gilbert, IEEE Trans. Magn. **40**, 3443 (2004).

[14] C. Dubs, O. Surzhenko, R. Thomas, J. Osten, T. Schneider, K. Lenz, J. Grenzer, R. Hübner, and E. Wendler, Phys. Rev. Mater. **4**, 024416 (2020).

[15] Q. Wang, B. Heinz, R. Verba, M. Kewenig, P. Pirro, M. Schneider, T. Meyer, B. Lägel, C. Dubs, T. Brächer, et al., Phys. Rev. Lett. **122**, 247202 (2019).

[16] B. Heinz, T. Brächer, M. Schneider, Q. Wang, B. Lägel, A. M. Friedel, D. Breitbach, S. Steinert, T. Meyer, M. Kewenig, C. Dubs, P. Pirro, and A. V. Chumak, Nano Lett. **20**, 4220 (2020).

[17] B. Heinz, Q. Wang, M. Schneider, E. Weiß, A. Lentfert, B. Lägel, T. Brächer, C. Dubs, O. V. Dobrovolskiy, P. Pirro, and A. V. Chumak, Appl. Phys. Lett. **118**, 132406 (2021).

[18] A. V. Chumak, in Spintronics Handbook: Spin Transport and Magnetism, 2nd ed (CRC Press, 2019), pp. 247–302. (1983)

[19] S. Wintz, V. Tiberkevich, M. Weigand, J. Raabe, J. Lindner, A. Erbe, A. Slavin, and J. Fassbender, Nature Nanotech. **11**, 948 (2016).

[20] H. Yu, O. d. Kelly, V. Cros, R. Bernard, P. Bortolotti, A. Anane, F. Brandl, F. Heimbach, and D. Grundler, Nat. Commun. **7**, 11255 (2016).

[21] P. Che, K. Baumgaertl, A. Kúkol'ová, C. Dubs, and D. Grundler, Nat. Commun. **11**, 1445 (2020).

[22] C. Liu, J. Chen, T. Liu, F. Heimbach, H. Yu, Y. Xiao, J. Hu, M. Liu, H. Chang, T. Stueckler, et al., Nat. Commun. **9**, 738 (2018).

[23] O. Dobrovolskiy, Q. Wang, D. Y. Vodolazov, B. Budinska, R. Sachser, A. Chumak, M. Huth, and A. Buzdin, arXiv:2103.10156.

[24] U. Garlando, Q. Wang, O. Dobrovolskiy, A. Chumak, and F. Riente, arXiv:2109.12973.

[25] P. Hansen, P. Röschmann, and W. Tolksdorf, J. Appl. Phys. **45**, 2728 (1974).

[26] P. Görnert and C. d'Ambly, Phys. Stat. Sol. (a) **29**, 95 (1975).

[27] J. Guigay, J. Baruchel, D. Challeton, J. Daval, and F. Mezei, J. Magn. Magn. Mater. **51**, 342 (1985).

[28] P. Röschmann, IEEE Trans. Magn. **17**, 2973 (1981).

[29] T. Sebastian, K. Schultheiss, B. Obry, B. Hillebrands, and H. Schultheiss, Front. Phys. **3**, 1589 (2015).

[30] J. J. Carmiggelt, O. C. Dreijer, C. Dubs, O. Surzhenko, and T. van der Sar, Appl. Phys. Lett. **119**, 202403 (2021).

[31] V. Bobkov and I. Zavislyak, Phys. Stat. Sol. (a) **164**, 791 (1997).

[32] B. D. Volkerts, Yttrium: Compounds, production and applications (Nova Science Publishers, Incorporated, 2011).

[33] P. Röschmann and W. Tolksdorf, Mater. Res. Bull. **18**, 449

[34] S. S. Kalarickal, P. Krivosik, M. Wu, C. E. Patton, M. L. Schneider, P. Kabos, T. J. Silva, and J. P. Nibarger, J. Appl. Phys. **99**, 093909(2006).

[35] B. Hillebrands, Rev. Sci. Instrum. **70**, 1589 (1999).

[36] B. A. Kalinikos, M. P. Kostylev, N. V. Kozhus, and A. N. Slavin, J. Phys.: Condens. Matter **2**, 9861 (1990).

[37] M. H. Seavey and P. E. Tannenwald, Phys. Rev. Lett. **1**, 168 (1958).

[38] B. A. Kalinikos and A. N. Slavin, J. Phys. C: Solid State Phys. **19**, 7013 (1986).

[39] S. Klingler, A. V. Chumak, T. Mewes, B. Khodadadi, C. Mewes, C. Dubs, O. Surzhenko, B. Hillebrands, and A. Conca, J. Phys. D: Appl. Phys. **48**, 015001 (2015).

# SUPPLEMENTAL MATERIALS: FMR CHARACTERIZATION

# Fast long-wavelength exchange spin waves in partially-compensated Ga:YIG

T. Böttcher,[1, 2] M. Ruhwedel,[1] K. O. Levchenko,[3] Q. Wang,[3] H. L. Chumak,[4] M. A. Popov,[4] I. V. Zavislyak,[4] C. Dubs,[5] O. Surzhenko,[5] B. Hillebrands,[1] A. V. Chumak,[3] and P. Pirro[1]

[1)]Fachbereich Physik and Landesforschungszentrum OPTIMAS, Technische Universität Kaiserslautern, Gottlieb-Daimler-Straße 46, 67663 Kaiserslautern, Germany
[2)]MAINZ Graduate School of Excellence, Staudingerweg 9, 55128 Mainz, Germany
[3)]Faculty of Physics, University of Vienna, A-1090 Wien, Austria
[4)]Faculty of Radiophysics, Electronics and Computer Systems, Taras Shevchenko National University of Kyiv, Kyiv, 01601, Ukraine
[5)]INNOVENT e.V. Technologieentwicklung, Prüssingstrasse 27B, 07745 Jena, Germany

Ferromagnetic Resonance – Vector Network Analyzer (FMR-VNA) spectroscopy is a fast and non-destructive technique that provides access to the fundamental magnetic properties of a material. This spectroscopy proved to be especially useful in the study of the magnetization in quaternary compounds of a partially compensated ferrimagnet Ga:YIG/GGG. By changing the concentration of Ga[S1-S4], it is possible to tune the saturation magnetization and the demagnetizing fields, opening a materials perspective towards magnonic logic devices[S5-S7] with isotropic spin-wave propagation[S8]. Considered as a milestone towards fully compensated antiferromagnetic Ga:YIG, it was discovered that the current samples already exhibit advantageous properties such as an induced perpendicular anisotropy.

Hence, the primary aim of this section is to establish a concise and precise interpretation of the ferromagnetic resonance data obtained for the thin epitaxial films of Ga:YIG/GGG and for the reference YIG/GGG film cut into 5 x 5 x 0.5 mm$^3$ pieces. To give our interpretation a better degree of flexibility, in the present analysis we have included samples grown by liquid phase epitaxy (LPE) on the substrates with different crystallographic directions – GGG (111) and GGG (001). The results presented in the sections below were obtained for several films: a 59 nm thick Ga:YIG/GGG(111) film, which was in the focus of interest for the BLS investigations in the main manuscript, a 105 nm thick Ga:YIG/GGG(111) film, a 95 nm thick Ga:YIG/GGG(001) film, and a reference 97 nm thick YIG/GGG(111) film. The Ga concentration in the investigated films is approximately $x_{Ga} \approx 1.0$, corresponding to saturation magnetization values of about 20 mT[S1], although the precise value is challenging to determine for such thin films on GGG substrate due to technical limitations.

FMR-VNA measurements were carried out in the frequency range of up to 20 GHz. To avoid non-linear contributions from magnon-magnon scattering processes that would contribute to the FMR linewidth broadening, we kept the RF power at 0 dBm. The measurement set-up consists of a VNA (Anritsu MS4642B) connected to an H-frame electromagnet GMW 3473-70 with an 8 cm air gap for various measurement configurations and magnet poles of 15 cm diameter to induce a sufficiently uniform biasing magnetic field. The electromagnet is powered by a bipolar power supply BPS-85-70-EC, allowing to generate $\approx 0.9$ T at 8 cm air gap. The calibrated VNA signal was transferred via SMA cables/non-magnetic SMA end-launch connectors to a straight Southwest Microwave RO4003.8mil microstrip with the sample mounted on top. All the measurements were performed at $T_r \approx 295\ K$. The measurements of the Ga:YIG samples proved to be challenging regarding lower applied magnetic fields. Therefore, to enhance the precision and the reliability of the results, an averaging procedure of the measured data was applied.

For the films grown on GGG(111), in-plane measurements (IP) were carried out with the magnetic field $H$ applied along the two orthogonal crystallographic axes $[11\bar{2}]$ and $[1\bar{1}0]$ corresponding to the sides of the samples, while for the film grown on GGG(001), $H$ was oriented along $[100]$ and $[1\bar{1}0]$ (along the side and diagonal of the rectangular samples). Out-of-plane (OOP) measurements were performed by applying the magnetic field normal to the surface of the films.

A model of the magnetic permeability tensor for the ferrites[S9] was adopted in order to determine the fundamental magnetic properties, such as the gyromagnetic ratio $\gamma$, the anisotropy fields $H_c$, $H_{u1}$, $H_{u2}$, and the effective magnetization $M_{\text{eff}}$. Furthermore, the Gilbert damping parameter $\alpha$ and the inhomogeneous linewidth broadening $\Delta H(0)$ were extracted from the FMR-VNA measurements.

## 1. MODEL USED TO ANALYZE THE DATA

For YIG films, a detailed description of the magnetocrystalline anisotropy (MCA) energy $U_A$ for different values of the magnetization together with a derivation of the specific anisotropy fields ansatz is found in the work of Bobkov and Zavislyak[S9].

The authors made an important conclusion regarding the influence of the substrate's crystallographic orientation and the direction of the applied magnetic field $H$. The work is mainly focused on a magnetostatic wave (MSW) analysis describing the frequency-dependent components of the dynamic permeability tensor $\mu$. The permeability tensor is derived in the linear approximation regarding the magnetization variables and includes the anisotropy constants up to the fourth-order terms.

We can adopt this approach to the IP FMR analysis using[S9]:

$$f^2 = \left(\frac{\gamma \mu_0}{2\pi}\right)^2 [(H_i + H_{A1} + M_s)(H_i + H_{A2}) - H_{A3}^2], \qquad (Eq.\ S1)$$

where $H_i = H_0$ – the internal magnetic field which is equal to the applied external field $H_0$ in this case, $\gamma = \frac{g\mu_B}{\hbar}$ – the gyromagnetic ratio with $g$ the g-factor, $\mu_B$ the Bohr magneton, and $\hbar$ the reduced Planck constant, $M_s$ – the saturation magentization. $H_{A1}$, $H_{A2}$, $H_{A3}$ are the combinations of the anisotropy terms derived from the magnetocrystalline (and magnetostrictive) energy density[S10] dependent on the magnetic field direction and the substrate's crystallographic orientations – see **Tables S1, S2**. For practical reasons, the original model was adjusted from the CGS to SI system of units, and the circular frequency $\omega$ was converted to the linear frequency $f = \omega/2\pi$.

The main mode for the OOP FMR is expressed as:

$$f^2 = \left(\frac{\gamma \mu_0}{2\pi}\right)^2 [(H_i + H_{A1})(H_i + H_{A2}) - H_{A3}^2], \qquad (Eq.\ S2)$$

where $\mu_0 H_i = \mu_0 H_0 - \mu_0 M_s$.

**Table S1.** Expressions for the $H_{A1}$, $H_{A2}$, $H_{A3}$ for ferrites on GGG(111) (adopted from [S9]).

|  | $H_{A1}$ | $H_{A2}$ | $H_{A3}$ |
| --- | --- | --- | --- |
| **IP**: $H \parallel [110]$ | $-\dfrac{K_c}{\mu_0 M_s} - \dfrac{2K_{u1}}{\mu_0 M_s}$ | $0$ | $-\sqrt{2}\dfrac{K_c}{\mu_0 M_s}$ |
| **OOP**: $H \parallel [001]$ | $-\dfrac{4}{3}\dfrac{K_c}{\mu_0 M_s} + \dfrac{2K_{u1}}{\mu_0 M_s} + \dfrac{4K_{u2}}{\mu_0 M_s}$ | | $0$ |

**Table S2.** Expressions for the $H_{A1}, H_{A2}, H_{A3}$ for ferrites on GGG(001) (adopted from [S9]).

|  | $H_{A1}$ | $H_{A2}$ | $H_{A3}$ |
|---|---|---|---|
| **IP**: $H \parallel [100]$ | $\dfrac{2K_c}{\mu_0 M_s} - \dfrac{2K_{u1}}{\mu_0 M_s}$ | $\dfrac{2K_c}{\mu_0 M_s}$ | 0 |
| **IP**: $H \parallel [110]$ | $\dfrac{K_c}{\mu_0 M_s} - \dfrac{2K_{u1}}{\mu_0 M_s}$ | $-\dfrac{2K_c}{\mu_0 M_s}$ | 0 |
| **OOP**: $H \parallel [001]$ | $\dfrac{2K_c}{\mu_0 M_s} + \dfrac{2K_{u1}}{\mu_0 M_s} + \dfrac{4K_{u2}}{\mu_0 M_s}$ |  | 0 |

The MCA of the YIG-based epitaxial films consists of the cubic ($U_A^{\text{cubic}}$), and the non-cubic part ($U_A^{\text{non cubic}}$). For the equilibrium magnetization under the linear approximation assumption, this leads to three major anisotropy fields contributing to an effective magnetization $\mu_0 M_{\text{eff}}$ – a cubic field $\left(H_c = \dfrac{K_c}{\mu_0 M_s}\right)$, a first order uniaxial field $\left(H_{u1} = \dfrac{2K_{u1}}{\mu_0 M_s}\right)$, and a second order uniaxial field $\left(H_{u2} = \dfrac{4K_{u2}}{\mu_0 M_s}\right)$[S9,S11]. The latter term, $H_{u2}$ appears only for the out-of-plane applied magnetic field, while the former two, $H_{u1}$ and $H_c$, contribute to the sample's magnetization in both IP and OOP configurations.

$K_c$ is the cubic anisotropy constant (denoted as $K_4$ in other sources[S12,S13]). $K_{u1}$ and $K_{u2}$ are the uniaxial out-of-plane anisotropies of first and second order, respectively. $K_{u1}$ can also be found as $K_{2\perp}$ in the literature[S12,S13], while $K_{u2}$ (and, consequently, $H_{u2}$) is a term, distinguished separately from $K_{u1}$ within the framework of the currently discussed model. Here, we define the cubic anisotropy as $H_c = \dfrac{K_c}{\mu_0 M_s}$, following the approach of Bobkov and Zavislyak[S9], however, in other articles[S12,S13] it is defined as $H_c^* = \dfrac{2 K_c}{\mu_0 M_s}$. However, the dependencies of the ferromagnetic resonance frequency on the cubic anisotropy are identical.

The cubic anisotropy originates from the preferred orientation of the magnetization along the crystallographic axis in the garnet lattice[S9, S12]. In the case of YIG, the negative $K_c$ has a positive contribution to the anisotropy field in (111) and a negative contribution in (001) films. For stoichiometric garnets grown under near-equilibrium conditions via the LPE technique, uniaxial anisotropy is mainly caused by the misfit strain between the lattice constants of the film and the substrate (see, e.g. S12 and Supplemental Materials of S12 for the strain calculations). This leads to a tensile stress and a stress-induced out-of-plane anisotropy contribution in the case of GGG substrates, dominating in Ga:YIG films over the weaker cubic and shape anisotropies.

It is also worth exploring the $H_{A3}$ anisotropy term more closely in the case of iron garnet/GGG (111) ferromagnetic resonance under IP magnetic field. For these films, as the [100] axis lies outside the film's plane, complementary measurements should be performed along the [11$\bar{2}$] direction. Based on Baselgia et al.[S15] $2H_c^2 = 2\left(\dfrac{K_4}{\mu_0 M_s}\right)^2 \cos^2(3\varphi)$, where $\varphi$ is the angle between the [1$\bar{1}$0] crystallographic axis, corresponding to the sample's side (see **Fig. S1**), and the magnetization $M$. For the $H \parallel [1\bar{1}0]$ case $\varphi = 0°$, $\cos^2(3\varphi) = 1 \rightarrow 2\left(\dfrac{K_4}{\mu_0 M_s}\right)^2 \cos^2(3\varphi) = 2H_c^2$, which holds true with a current model. However, for the $H \parallel [11\bar{2}]$ case $\varphi = 90°$, $\cos^2(3\varphi) = 0 \rightarrow 2\left(\dfrac{K_4}{M}\right)^2 \cos^2(3\varphi) = 0$ and the $2H_c^2$ contribution vanishes. This simple dependency allows to quickly express the cubic field $H_c$ for iron garnet/GGG(111) through the measurements of two well-defined directions, corresponding in our case to the sample's sides, opposing to the more demanding angle-resolved measurements.

## 2. IRON GARNET/GGG(111)

Let us consider thin YIG and GaYIG films grown on a GGG(111) substrate (**Fig. S1**). Then, based on (Eq. S1, S2) with substituted corresponding anisotropy terms from **Table S1, S2** we obtain equations for the FMR frequency $f_\parallel$ under the IP magnetic field $H^\parallel$:

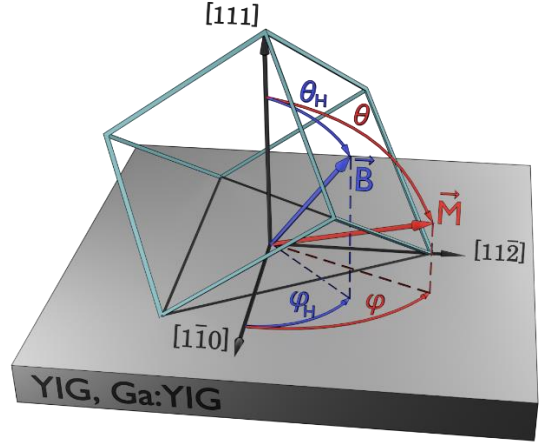

**Figure S1.** Crystallographic orientations of YIG/GGG(111) and Ga:YIG/GGG(111).

$$f_\parallel [1\bar{1}0] = \frac{\gamma\mu_0}{2\pi}\sqrt{H^\parallel(H^\parallel - H_c - H_{u1} + M_s) - 2H_c^2} \qquad \text{(Eq. S3)}$$

$$f_\parallel [11\bar{2}] = \frac{\gamma\mu_0}{2\pi}\sqrt{H^\parallel(H^\parallel - H_c - H_{u1} + M_s)} \qquad \text{(Eq. S4)}$$

$$f_\perp = \frac{\gamma\mu_0}{2\pi}(H^\perp - \frac{4}{3}H_c + H_{u1} + H_{u2} - M_s), \qquad \text{(Eq. S5)}$$

The $2H_c^2$ term is usually very small compared to the product before it. For currently investigated Ga:YIG film, it introduces an approximate shift in the FMR frequency (rounded to the highest value to incorporate a measurement error) of $\Delta f_\parallel \approx 110$ MHz for $\mu_0 H^\parallel \approx 110$ mT $(@f_\parallel \approx 2$ GHz$)$ and $\Delta f_\parallel \approx 3$ MHz for $\mu_0 H^\parallel \approx 700$ mT $(@f_\parallel \approx 19.5$ GHz$)$. Therefore, the term $2H_c^2$ can be neglected for frequencies above 5 GHz as it is done in the analysis of the BLS data in the current manuscript.

Similar results with slightly different approximations are presented in a recent work of Dubs et al.[S12], where sub-100 nm YIG films were analyzed via the angle-resolved broadband FMR-VNA. There, only results for samples grown on (111) substrates are shown and the main focus was made on the influence of the YIG film thickness on static and dynamic magnetic properties. The main difference between the theoretical model used in the aforementioned publication and our approach stems from a slightly different interpretation of the uniaxial / stress-induced anisotropy. However, assuming that $H_c = \frac{K_c}{\mu_0 M_s} = \frac{K_4}{\mu_0 M_s}$, $H_{u1} = \frac{2K_{u1}}{\mu_0 M_s} = H_{2\perp} - H_c$, adding the second-order uniaxial anisotropy, and omitting the relatively weak in-plane anisotropy field $\frac{2K_{u\parallel}}{\mu_0 M_s}$, we will get to the same set of equations describing the ferromagnetic resonance in both the IP and OOP configurations.

A thorough investigation of the static and dynamic magnetic properties of Ga:YIG(111) was performed in a recent work of Joris J. Carmiggelt et al.[S13]. The authors have also underlined the role of anisotropy in switching the easy magnetization axis to OOP, as seen in the FMR measurements performed on their 45 nm thick sample. Precise measurements in the IP and OOP FMR configurations were performed, and the accumulated data was fit with the respective Kittel equations, modified to incorporate both the cubic and the uniaxial anisotropy. Similar to the earlier discussed work of Dubs et al.[S12], the authors do not consider the uniaxial anisotropy of the second order for OOP resonance, but analytically determine the uniaxial out-of-plane anisotropy $\frac{2K_{2\perp}}{\mu_0 M_s} = H_{2\perp}$, which includes both the cubic and the stress-induced anisotropy contributions. Considering different thicknesses and slightly different $x_{Ga}$ concentrations in Carmiggelt's work, the expected slight change in cubic ($-4.1$ mT) and uniaxial anisotropy ($104.7$ mT) is observed[S13].

Otherwise, both models are in good agreement and lead to similar results.

## 2.1. Determining $H_c$ from FMR

Considering the many unknown parameters $\gamma, M_s, H_c, H_{u1},$ and $H_{u2}$ in the Equations S3 – S5, the fitting will yield high error margins and poor convergence. Hence, a more elaborated treatment is required to specify the terms.

To determine the cubic anisotropy, samples are measured while magnetized along the $[1\bar{1}0]$ and the $[11\bar{2}]$ crystallographic directions. The applied in-plane magnetic field was selected in a specific range higher than the saturation field $H_s$ (at which the sample is getting homogenously magnetized in-plane) but still low enough to deduce the $2H_c^2$ term. The saturation field is discussed in the next section and can also be extracted from **Fig. S2**.

Subtracting Equations S4 and S3, we obtain:

$$H_c = \sqrt{\frac{1}{2}\left(\frac{2\pi f_{\parallel [11\bar{2}]}}{\gamma \mu_0}\right)^2 - \frac{1}{2}\left(\frac{2\pi f_{\parallel [1\bar{1}0]}}{\gamma \mu_0}\right)^2} \qquad \text{(Eq. S6)}$$

Usually, the cubic anisotropy field $\mu_0 H_c$ varies from $\approx -4..5$ mT for micrometer-thick and bulk YIG[S16] or about $-4.2$ mT for sub-40 nm LPE YIG films[S12]. Evaluation of our experimental data using (Eq. S6), gives the values $\mu_0 H_c \approx -5.05 \pm 0.5$ mT for 97 nm thick YIG film, $-4.2 \pm 0.7$ mT and $-6.5 \pm 1.3$ mT for 105 nm and 59 nm thick Ga:YIG samples correspondingly. This is in good agreement with $-4.1$ mT found for the 45 nm Ga:YIG film[S13].

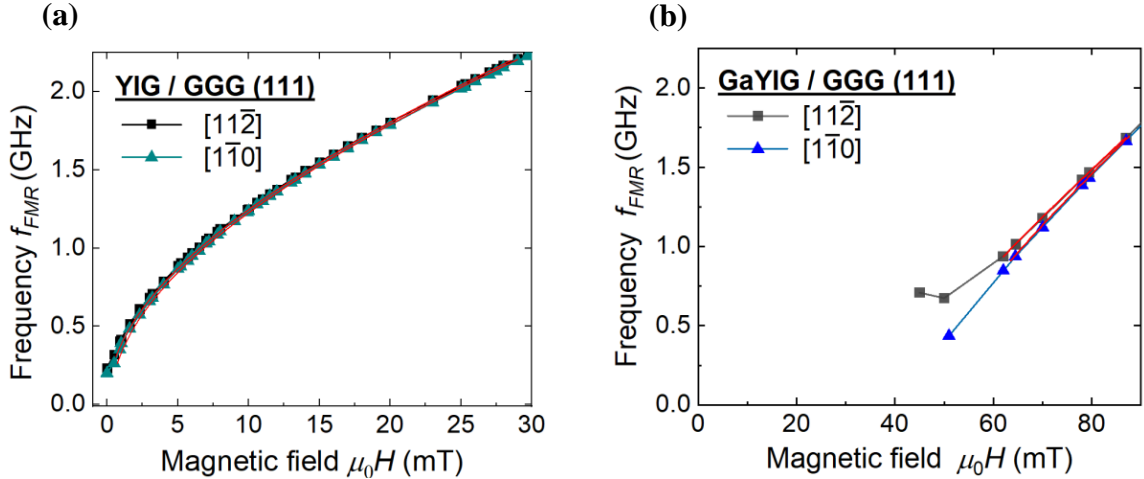

**Figure S2.** Cubic anisotropy $H_c$ fitting for the 97 nm YIG (a) and 105 nm Ga:YIG (b) films grown on GGG (111). For Ga:YIG @ $\mu_0 H \approx$ 50 mT, a kink in the low field regime in (b) indicates a transition from unsaturated (multidomain state) to a homogeneous magnetization, while for YIG such a transition is barely visible (in-plane bias field).

## 2.2. Determining $H_{u1}$ from FMR

The uniaxial anisotropy of the first order can be fitted directly to the measured data for the in-plane resonance (Eq. S3-S4) if the saturation magnetization $\mu_0 M_s$ is known. In the absence of this value, it is better to use complementary equations to increase the fitting reliability and specify the range in which to search for $H_{u1}$. Here, we propose an equation[S9,S11,S17], that includes anisotropy fields relevant for the in-plane configuration:

$$\mu_0^2 H_s (H_s - H_c - H_{u1}) - 2(\mu_0 H_c)^2 = 0 \qquad \text{(Eq. S7)}$$

where $\mu_0 H_s$ is the saturation field required to rotate the magnetization of the film in plane by suppressing the strong perpendicular anisotropy in Ga:YIG[S13]. Equation S7, was derived for iron garnet/GGG(111), $f @ H \parallel [1\bar{1}0]$ assuming $\mu_0 H = \mu_0 H_s$ and $f = 0$[S9].

The unsaturated region is identified on the $f_{FMR}(H^{\|})$ plot as an inverse dependence of the FMR frequency on the applied magnetic field. The critical field $H_s$ is indicated by the lowest extremum on this dependence[S13]. The signal below the field $H_s$ is hardly distinguishable from the noise background. Typical values for the micrometer thick YIG are around 5 mT. For the 45-nm thick Ga:YIG film, the magnetization reaches saturation at about 87 mT[S13].

In our measurements, the saturation field $\mu_0 H_s$ of Ga:YIG/GGG(111) was about $50 \pm 5$ mT for the 105 nm film (see **Fig. S2 b**) and about $81 \pm 2$ mT for the 59 nm film. An alternative Ga:YIG sample of a relatively similar thickness (56 nm), matching $M_s$ (20.2 mT), and grown via the LPE under the same conditions clearly shows (**Fig. S3 a**) a negative derivative of the frequency with respect to the magnetic field, $\frac{\partial f}{\partial H} < 0$, until $\mu_0 H_s \cong 78 \pm 2$ mT. Above this field, the resonances are more pronounced and the derivative is positive, $\frac{\partial f}{\partial H} > 0$ (**Fig. S3 a**). The 59 nm thick Ga:YIG film had shown the same behavior and reached saturation at a similar field $\mu_0 H_s \cong 81 \pm 2$ mT, corresponding to the resonance in a slightly lower frequency $f_{FMR} \cong 1.0$ GHz. For YIG/GGG(111) the transition was below 2 mT, and was hard to define.

Having both $H_s$ and $H_c$ fields, we may derive $H_{u1}$ from (Eq. S7) and increase precision of the obtained value through the fit (Eq. S3-S4). Uniaxial anisotropy was estimated to be around $74.8 \pm 1$ mT for the 105 nm thick Ga:YIG and around $94.1 \pm 0.5$ mT for 59 nm thick Ga:YIG films.

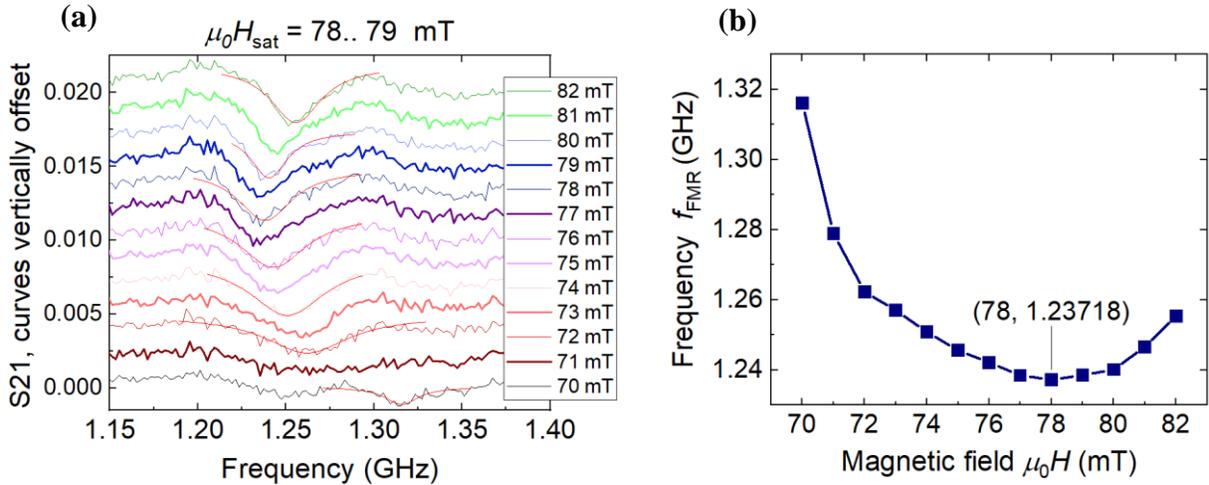

**Figure S3.** Magnetization saturation of the 56-nm Ga:YIG/GGG(111) thick film. **(a)** Color map of the data collected from the $S_{21}$ VNA trace as a function of the in-plane magnetic field, $H^{\|} \| [11\bar{2}]$, increasing with a step $\Delta \mu_0 H = 1$ mT. **(b)** Ferromagnetic resonance $f_{FMR}$ as a function of the applied magnetic field obtained from the fit of the corresponding resonance curves in **(a)** with a Lorentzian.

### 2.3. Determining $\gamma$, $M_s$, and $H_{u2}$ from FMR

The saturation magnetization $M_s$ can be derived from the FMR measurements using (Eq. S3 – S4) if the fields $H_c$ and $H_{u1}$ are known.

In the case of pure YIG, anisotropy contributions $H_{u1}$ and $H_c$ are usually two orders of magnitude smaller than the saturation magnetization $M_s$. Hence, these two terms are often either neglected or combined in one term[S18]. However, they start to play a crucial role in Ga:YIG films, leading to the negative effective magnetization $\mu_0 M_{eff} = \mu_0 (M_s - H_c - H_{u1})$ (**Fig. S4**) as determined from the direct Kittel fitting of $f_{FMR}(H^{\|})$. One of the reasons behind the negative values of $M_{eff}$ in Ga:YIG is a substitution of the magnetic $Fe^{3+}$ ions with the non-magnetic $Ga^{3+}$ ions in a tetrahedral sub-lattice leading to a decreased $M_s$. The second reason is the pronounced uniaxial anisotropy $H_{u1}$ of the Ga:YIG films due to a large lattice misfit strain between the film and the GGG substrate.

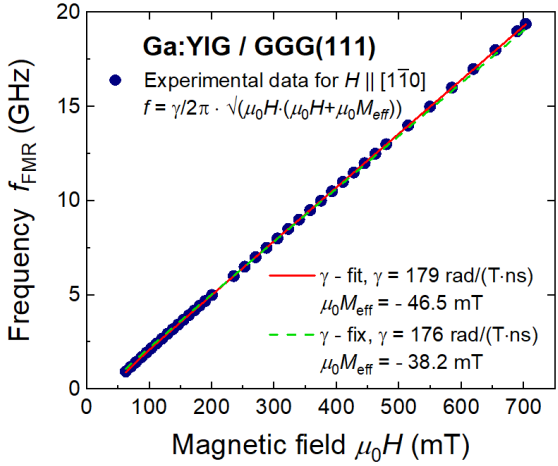

**Figure S4.** Simplified Kittel equation fit (red and green dashed lines) to 105 nm thick Ga:YIG/GGG (111) experimental FMR-VNA data (blue circles). The red line corresponds to the fitted gyromagnetic ratio $\gamma$, while the green dashed line – to the fixed value $\gamma = 176\ \frac{\text{rad}}{\text{ns}\cdot\text{T}}$. In both assumptions, the term $\mu_0 M_{\text{eff}}$ yields a negative value hinting a decreased saturation magnetization and a strong anisotropy contribution.

Domination of the uniaxial anisotropy over the saturation magnetization in Ga-substituted YIG samples suggests a perpendicular magnetic anisotropy with the easy magnetization pointing out of plane.

The saturation magnetization $M_s$ determined using FMR was compared to the values obtained using vibrational sample magnetometry (VSM). The values differ only by 2% for the 105 nm-thick and by 19.8 % for the 59 nm-thick Ga:YIG films. Moreover, the described methodology does not allow for the determination of $M_s$ for the films grown on (001) GGG substrates. In the following, we use the VSM values to determine the saturation magnetization $M_s$, and the FMR-VNA spectroscopy to define all the anisotropy fields contributions.

The gyromagnetic ratio $\gamma$ could be fixed to the free-electron value $176\ \frac{\text{rad}}{\text{T}\cdot\text{ns}}$, as it is typically done for the YIG films, or could be fitted experimentally. The difference in the results for the effective magnetization $M_{\text{eff}}$ was reaching quite a substantial value of 20 % – see **Fig. S4**. The obtained $\gamma$ values are given in **Table 3**, and are around $179\ \frac{\text{rad}}{\text{T}\cdot\text{ns}}$ for Ga:YIG and $177\ \frac{\text{rad}}{\text{T}\cdot\text{ns}}$ for YIG.

The uniaxial anisotropy of the second order, $\mu_0 H_{u2}$, is obtained through the OOP FMR measurements and the subsequent fitting with Equation S5.

The parameters obtained following the described procedure are given in **Tables S3**.

## 3. IRON GARNET/GGG(001)

The crystallographic orientation (**Fig. S5**) of the substrate influences the anisotropy ansatz (see **Table S2**). For Ga:YIG/GGG(001), the in-plane and out-of-plane ferromagnetic resonances are described through a system of Equations S8 – S10.

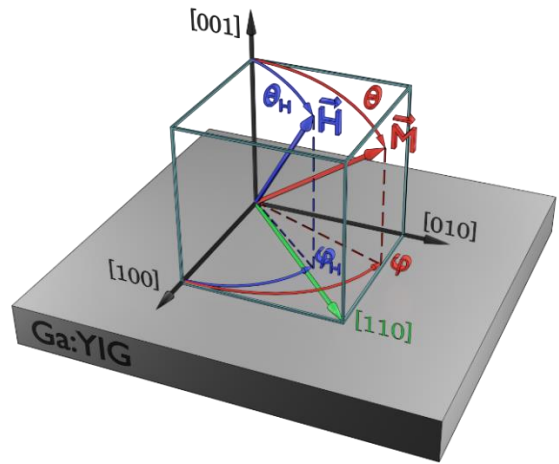

**Figure S5.** Crystallographic orientations of Ga:YIG/GGG(001).

$$f_{||}[1\bar{1}0] = \frac{\gamma\mu_0}{2\pi}\sqrt{(H^{||} - 2H_c)(H^{||} + H_c - H_{u1} + M_s)} \qquad \text{(Eq. S8)}$$

$$f_{||}[100] = \frac{\gamma\mu_0}{2\pi}\sqrt{(H^{||} + 2H_c)(H^{||} + 2H_c - H_{u1} + M_s)} \qquad \text{(Eq. S9)}$$

$$f^{\perp} = \frac{\gamma\mu_0}{2\pi}(H^{\perp} + H_{u1} + 2H_c + H_{u2} - M_s) \qquad \text{(Eq. S10)}$$

The case of (001) crystallographic anisotropy is more complex to analyze compared to the (111) case since (1) it is not possible to introduce the same effective magnetization $M_{\text{eff}}$ for IP and OOP configurations, and (2) the algorithm developed for the FMR extraction of the saturation magnetization $M_s$ is not applicable. The VSM values for $M_s$ are used in the following.

### 3.1 Determining $\gamma, H_c, H_{u1}, H_{u2}$ from FMR

Since the cubic anisotropy field $H_c$ could not be separately expressed like in the case of (111) films, $H_c$ was fitted simultaneously with $H_{u1}$ and $\gamma$ for IP measurements with (Eq. S8-S10). If there is a need to increase the precision of the fitting, the complementary equations derived from the saturation magnetization under the assumptions $H^{||} = H_s$, $f = 0$ can be used:

$$\begin{cases} H || [1\bar{1}0]: \mu_0^2(H_s - 2H_c)(H_s + H_c - H_{u1}) = 0 \\ H || [100]: \mu_0^2(H_s + 2H_c)(H_s + 2H_c - H_{u1}) = 0 \end{cases} \qquad \text{(Eq. S11)}$$

An important difference in the measurement approach for the sample on GGG(001) is based upon a different set of IP crystallographic axes, $[1\bar{1}0]$ (diagonal) and $[100]$ (side), along which the magnetic field $H^{||}$ is applied.

Conclusively, the parameters obtained according to the procedures described above are summarized in **Table S3** for all the samples under the investigation.

**Table S3.** The parameters obtained from FMR-VNA analyses for YIG/GGG (111), Ga:YIG/GGG(111) and Ga:YIG/GGG(001) films. $\mu_0 M_s$ (Ga:YIG) obtained from VSM.

| Garnet/ substrate | Thickness, $t$ | Gyromagnetic ratio, $\gamma$ rad/ns·T | $\mu_0 M_s$ mT | $\mu_0 H_c$ mT | $\mu_0 H_{u1}$ mT | $\mu_0 H_{u2}$ mT |
|---|---|---|---|---|---|---|
| YIG GGG (111) | 97 nm | **Fit: 177** | **Fit: 182.4** $\pm$ 1.8 | $-5.1$ $\pm$ 0.5 | $-3.5$ $\pm$ 0.5 | 3.6 $\pm$ 0.5 |
| Ga:YIG GGG (111) | 59 nm | **Fit: 179** | **Fix: 20.2** | $-6.5$ $\pm$ 1.3 | 94.1 $\pm$ 0.5 | 2.4 $\pm$ 0.1 |
| Ga:YIG GGG (111) | 105 nm | **Fit: 179** | **Fix: 24.4** | $-4.2$ $\pm$ 0.7 | 74.8 $\pm$ 1.0 | 2.1 $\pm$ 0.1 |
| Ga:YIG GGG (001) | 96 nm | **Fit: 179** | **Fix: 21.7** | $-5.3$ $\pm$ 0.6 | 92.4 $\pm$ 1.7 | 5.2 $\pm$ 0.2 |

## 4. DETERMINING $\alpha$, $\Delta H(0)$ FROM FMR

The FMR linewidth $\Delta H$ depends on the ferromagnetic resonance $f_{FMR}(H^{\|})$ frequency according to[S19,]:

$$\mu_0 \Delta H = \mu_0 \Delta H(0) + \frac{\alpha \, 4\pi \, f_{FMR}}{\gamma}, \tag{Eq. S12}$$

where $\mu_0 \Delta H$ is the FMR full width at half maximum (FWHM), $\mu_0 \Delta H(0)$ – the inhomogeneous linewidth broadening, $\alpha$ – the Gilbert damping parameter, and $\mu_0$ – the permeability of free space.

In order to recalculate $\Delta f$, obtained from the broadband frequency FMR-VNA measurements, into $\Delta H$ in (Eq. S12), one can use the approach introduced by Kalarickal et al.:

$$2\pi \, \Delta f = \Delta H \, \frac{\partial f_{Kittel}(H^{\|})}{\partial H^{\|}}\bigg|_{H^{\|} = H_{Kittel}(f_{FMR})} = \mu_0 \, \Delta H \, \gamma \, P_A(f_{FMR}) \tag{Eq. S13}$$

$$\mu_0 \Delta H = \frac{2\pi \, \Delta f}{\gamma \, P_A(f_{FMR})}, \tag{Eq. S14}$$

where $P_A(f_{FMR}) = \sqrt{1 + \left(\frac{\gamma \, \mu_0 \, M_s}{4\pi \, f_{FMR}}\right)^2}$. The term $2\pi$ was included in the Equation S14 to recalculate the linear frequency $f$ from the angular frequency $\omega$.

The original formulas (Eq. S12-S14) were derived for YIG/GGG(111) considering $M_s \gg |H_{u1}|, |H_c|$. Hence, in the differentiated ferromagnetic resonance equation $\frac{\partial f_{Kittel}(H^{\|})}{\partial H^{\|}}$, the effective magnetization $M_{eff}$ was substituted with the saturation magnetization $M_s$. However, for the Ga:YIG films, as shown earlier in this section, $M_s < |H_{u1}|$, and the direction of the applied magnetic field $H^{\|}$ with respect to the specific GGG substrate influences the resonance equation. Therefore, to obtain an appropriate field swept linewidth, each of the specific ferromagnetic resonance equations (Eq. S3-S4, Eq. S8-S9) should be differentiated separately with the anisotropy fields included (**Table S4**).

Here, we discuss the results only for the in-plane configuration, as the damping constant $\alpha$ is enhanced in the out-of-plane measurements. This is attributed to the influence of a magnetically inhomogeneous transient layer near the substrate interface[S12].

**Table S4.** The expressions for the recalculated field linewidth $\mu_0 \Delta H$ from the frequency linewidth $\Delta f$ based upon the magnetic field orientation with respect to the crystallographic axis of the GGG substrate.

| Substrate | Direction of $\mu_0 H^{\|}$ | $\mu_0 \Delta H$ (recalculated from $\Delta f$) |
|---|---|---|
| GGG (111) | $\mu_0 H^{\|} \, \| \, [1\bar{1}0]$ | $\dfrac{8\pi^2 \, \Delta f \cdot f_{FMR[1\bar{1}0]}}{\mu_0 \gamma^2 \, (2H^{\|} - H_c - H_{u1} + M_s)}$ |
| | $\mu_0 H^{\|} \, \| \, [11\bar{2}]$ | $\dfrac{8\pi^2 \, \Delta f \cdot f_{FMR[11\bar{2}]}}{\mu_0 \gamma^2 \, (2H^{\|} - H_c - H_{u1} + M_s)}$ |
| GGG (001) | $\mu_0 H^{\|} \, \| \, [1\bar{1}0]$ | $\dfrac{8\pi^2 \, \Delta f \cdot f_{FMR[1\bar{1}0]}}{\mu_0 \gamma^2 \, (2H^{\|} - H_c - H_{u1} + M_s)}$ |
| | $\mu_0 H^{\|} \, \| \, [100]$ | $\dfrac{8\pi^2 \, \Delta f \cdot f_{FMR[100]}}{\mu_0 \gamma^2 \, (2H^{\|} + 4H_c - H_{u1} + M_s)}$ |

In a specific case for the iron garnet/GGG (111) films under the in-plane magnetic field $\mu_0 H^{\parallel} \parallel [11\bar{2}]$ or under $\mu_0 H^{\parallel} \parallel [1\bar{1}0]$ field in the frequency range above 5 GHz, it is possible to derive $\mu_0 \Delta H$ similar to Kalarickal et al.. Because the anisotropy fields ensemble in the ferromagnetic resonance equation (Eq. S4) assumes plain form $\mu_0(-H_c - H_{u1} + M_s) = \mu_0 M_{eff}$, the expression given in the second row in **Table S4** could be re-written as:

$$\frac{2\pi \Delta f}{\gamma \sqrt{1+\left(\frac{\gamma \mu_0 M_{eff}}{4\pi f_{FMR[11\bar{2}]}}\right)^2}} \quad \text{(Eq. S15)}$$

Based on the (Eq. S12) and expressions from **Table S4**, the Gilbert damping constant $\alpha$ and the inhomogeneous linewidth broadening $\mu_0 \Delta H(0)$ were calculated, and, subsequently, summarized in **Table S5**. The errors were calculated based on the corresponding fits convergences.

**Table S5.** The damping parameters obtained from the FMR-VNA analyses for the thin films.

| Iron garnet/ substrate | Thickness, $t$ (nm) | $\alpha$, $10^{-4}$ | | $\mu_0 \Delta H(0)$ | | $\mu_0 \Delta H$ (mT) @ $f \approx 10.5$ GHz | | $\Delta f$ (MHz) @ $f \approx 10.5$ GHz | |
|---|---|---|---|---|---|---|---|---|---|
| GGG (111) | | $[11\bar{2}]$ | $[1\bar{1}0]$ | $[11\bar{2}]$ | $[1\bar{1}0]$ | $[11\bar{2}]$ | $[1\bar{1}0]$ | $[11\bar{2}]$ | $[1\bar{1}0]$ |
| YIG | 97 nm | 1.3 ±0.15 | 0.6 ±0.16 | 0.1 ±0.01 | 0.2 ±0.01 | 0.195 ±0.003 | 0.313 ±0.008 | 5.7 ±0.1 | 9.1 ±0.23 |
| Ga:YIG | 59 nm | 6.1 ±0.62 | 4.3 ±1.02 | 0.4 ±0.05 | 0.7 ±0.08 | 0.786 ±0.017 | 0.934 ±0.024 | 22.5 ±0.49 | 26.9 ±0.69 |
| Ga:YIG | 105 nm | 4.6 ±0.28 | 4.9 ±0.52 | 0.4 ±0.02 | 0.4 ±0.04 | 0.637 ±0.009 | 0.727 ±0.017 | 18.2 ±0.27 | 20.8 ±0.50 |
| GGG(001) | | [100] | $[1\bar{1}0]$ | [100] | $[1\bar{1}0]$ | [100] | $[1\bar{1}0]$ | [100] | $[1\bar{1}0]$ |
| Ga:YIG | 96 nm | 8.4 ±0.85 | 6.7 ±0.68 | 0.4 ±0.06 | 0.4 ±0.05 | 1.066 ±0.055 | 0.804 ±0.037 | 30.9 ±1.60 | 23.4 ±1.06 |

Divergence between the Gilbert damping constant $\alpha$ along the different crystallographic directions hints a pronounced influence of the inhomogeneous linewidth broadening $\mu_0 \Delta H(0)$, but might be also associated with a relatively large error bar. Therefore, a more detailed investigation is required to verify the origins if this phenomenon.

To compare these values with the literature, it is worth to mention that the typical values of Gilbert damping parameter $\alpha$ of discs made from the bulk crystals[S21] are $\alpha = 0.4 \cdot 10^{-4}$ for YIG and $\alpha = 1.25 \cdot 10^{-4}..\ 2.44 \cdot 10^{-4}$ for Ga:YIG ($x_{Ga} = 0.78..0.88$) [respectively[S22,S21]]. Epitaxially-grown micrometer-thick YIG LPE films have slightly higher damping. Their Gilbert parameters range from $\alpha = 0.4 \cdot 10^{-4}$ ($t = 23$ μm)[S21] to $0.5 \cdot 10^{-4}$ ($t = 3$ μm)[S14]. High-quality LPE YIG films with thickness down to hundreds of nanometers are reported to possess $\alpha = 1.0..2.0 \cdot 10^{-4}$ ($t = 200$ nm)[S23,S24], $\alpha = 1.7 \cdot 10^{-4}$ ($t = 100$ nm)[S14]. A new dimensionality milestone was achieved with the high-quality sub-100 nm LPE-grown YIG films[S12], that were shown to exhibit low ferromagnetic losses and relatively low Gilbert damping $\alpha = 1.0 \cdot 10^{-4}..\ 1.2 \cdot 10^{-4}$ ($t = 42..11$ nm)[S12]. Just recently, a 45 nm thick Ga:YIG film was reported to have $\alpha = 1.0 \cdot 10^{-3}$ [S13], which is higher compared to the $\alpha = 6.1 \cdot 10^{-4}$ presented in this study. However, considering thinner sample with a slightly lower saturation magnetization in the work of Carmiggelt et al.[S13], both damping constants are in relatively good agreement.

**LITERATURE**:

[S1] P. Hansen, P. Röschmann, W. Tolksdorf "Saturation magnetization of gallium-substituted yttrium iron garnet", *J. Appl. Phys.* **45**, 2728-27-32 (1974). DOI: 10.1063/1.1663657

[S2] P. Görnert and C. d'Ambly "Investigations of the growth and the saturation magnetization of garnet single crystals $Y_3Fe_{5-x}Ga_xO_{12}$ and $Y_3Fe_{5-x}Al_xO_{12}$", *PSS* (a) **29** (1975). DOI: 10.1002/pssa.2210290111

[S3] J. Guigay, J. Baruchel, D. Challeton, J. Daval and F. Mezei "Local measurement of magnetization in two Ga-YIG single crystals grown by different methods", *J. Magn. Magn. Mater* **51**, 342 (1985). DOI: 10.1016/0304-8853(85)90034-4

[S4] P. Röschmann "Annealing effects on FMR linewidth in Ga substituted YIG", *IEEE Transactions on Magnetics* **17**(6), 2973 (1981). DOI: 10.1109/TMAG.1981.1061632

[S5] A. Mahmoud, F. Ciubotaru, F. Vanderveken, A. V. Chumak, S. Hamdioui, C. Adelmann and S. Cotofana "Introduction to spin wave computing (Tutorial Article)", *J. of Appl. Phys.* **128**, 161101 (2020). DOI: 10.1063/5.0019328

[S6] P. Pirro, V. I. Vasyuchka, A. A. Serga, et al., "Advances in coherent magnonics", *Nat. Rev. Mater.* (2021). DOI: 10.1038/s41578-021-00332-w

[S7] A. V. Chumak, et al., "*Roadmap on spin-wave computing*", (2021). arXiv:2111.00365

[S8] S. Klingler, P. Pirro, T. Brächer, B. Leven, B. Hillebrands, A. V. Chumak "Spin-wave logic devices based on isotropic forward volume magnetostatic waves", *Appl. Phys. Lett.*, **106**(2), 212406 (2015). DOI: 10.1063/1.4921850

[S9] V. B. Bobkov, I. V. Zavislyak "Equilibrium State and Magnetic Permeability Tensor of the Epitaxial Ferrite Films", *Phys. Stat. Sol.* (a) **164**, 791 (1997). DOI: 10.1002/1521-396X(199712)164:2<791::AID-PSSA791>3.0.CO;2-7

[S10] H. Szymczak and N. Tsuya "Phenomenological Theory of Magnetostriction and Growth-Induced Anisotropy in Garnet Films", *Phys. Stat. Sol.* (a) **54**, 117 (1979). DOI: 10.1002/pssa.2210540115

[S11] I. V. Zavislyak and M. A. Popov "Yttrium: Compounds, Production and Applications" edited by B. D. Volkerts, **Chapter 3**, Nova Science Publishers, (2009).

[S12] C. Dubs, O. Surzhenko, R. Thomas, J. Osten, T. Schneider, K. Lenz, J. Grenzer, R. Hübner, and E. Wendler "Low damping and microstructural perfection of sub-40nm-thin yttrium iron garnet films grown by liquid phase epitaxy", *Phys. Rev. Materials* **4**, 024416 (2020). DOI: 10.1103/PhysRevMaterials.4.024416

[S13] J. J. Carmiggelt, O.C. Dreijer, C. Dubs, O.Surzhenko, T. van der Sar "Electrical spectroscopy of the spin-wave dispersion and bistability in gallium-doped yttrium iron garnet", *Appl. Phys. Lett.* **119**, 202403 (2021). DOI: 10.1063/5.0070796

[S14] C. Dubs, O. Surzhenko, R. Linke, A. Danilewsky, U. Brckner, and J. Dellith "Sub-micrometer yttrium iron garnet LPE films with low ferromagnetic resonance losses", *J. Phys. D: Appl. Phys.*, 50(**20**), 204005 (2017). DOI: 10.1088/1361-6463/aa6b1c

[S15] L. Baselgia, M. Warden, F. Waldner, Stuart L. Hutton, John E. Drumheller, Y. Q. He, P. E. Wigen, and M. Maryško, "Derivation of the resonance frequency from the free energy of ferromagnets", *Phys. Rev. B* **38**, 2237 (1988). DOI: 10.1103/PhysRevB.38.2237

[S16] S. A. Manuilov, S. I. Khartsev, and A. M. Grishin "Pulsed laser deposited $Y_3Fe_5O_{12}$ films: Nature of magnetic anisotropy I", *J. Appl. Phys.* **106**, 123917 (2009) DOI: 10.1063/1.3272731

[S17] I. V. Zavislyak, M. A. Popov, G. Sreenivasulu, and G. Srinivasan "Electric field tuning of domain magnetic resonances in yttrium iron garnet films", *Appl. Phys. Lett.* **102**, 222407 (2013). DOI: 10.1063/1.4809580

[S18] M. C. Onbasli, A. Kehlberger, D. H. Kim, G. Jakob, M. Kläui, A. V. Chumak, B. Hillebrands, and C. A. Ross "Pulsed laser deposition of epitaxial yttrium iron garnet films with low Gilbert damping and bulk-like magnetization", *APL Mater.* **2**, (2014). DOI: 10.1063/1.4896936

[S19] P. Pirro, T.Brächer, A. V. Chumak, B. Lägel, C. Dubs, O. Surzhenko, P. Görnert, B. Leven, and B. Hillebrands "Spin-wave excitation and propagation in microstructured waveguides of yttrium iron garnet/Pt bilayers", *Appl. Phys. Lett.* **104**, 012402 (2014). DOI: 10.1063/1.4861343

[S20] S. S. Kalarickal, P. Krivosik, M. Wu, C. E. Patton, M. L. Schneider, P. Kabos, T. J. Silva, and J. P. Nibarger "Ferromagnetic resonance linewidth in metallic thin films: Comparison of measurement methods", *J. Appl. Phys.* **99**, 093909 (2006). DOI: 10.1063/1.2197087

[S21] P. Röschmann and W. Tolksdorf "Epitaxial growth and annealing control of FMR properties of thick homogeneous Ga substituted yttrium iron garnet films", *Mat. Res. Bull.* **18**, 449 (1983). DOI: 10.1016/0025-5408(83)90137-X

[S22] P. Röschmann "Annealing effects on FMR linewidth in Ga substituted YIG", in *IEEE Transactions on Magnetics*, **17** (6), 2973, (1981). DOI: 10.1109/TMAG.1981.1061632

[S23] S. Maendl, I. Stasinopoulos, D. Grundler "Spin waves with large decay length and few 100 nm wavelengths in thin yttrium iron garnet grown at the wafer scale", *APL* **111**, 012403 (2017). DOI: 10.1063/1.4991520

[S24] C. Hahn, G. de Loubens, O. Klein, M. Viret, V. V. Naletov, and J. Ben Youssef "Comparative measurements of inverse spin Hall effects and magnetoresistance in YIG/Pt and YIG/Ta", *Phys. Rev. B* **87**(17), 174417 (2013). DOI: 10.1103/PhysRevB.87.174417